\documentclass[prd,nofootinbib]{revtex4}
\usepackage{graphicx}
\usepackage{amsfonts}
\usepackage{latexsym}
\usepackage{amssymb}
\usepackage{epsfig}
\usepackage{amsmath,amssymb}

\begin{document}
 
\title{
 Thermal leptogenesis 
  in a supersymmetric neutrinophilic Higgs model
}

\author{Naoyuki Haba}
 \email{haba@phys.sci.osaka-u.ac.jp}
 \affiliation{
 Department of Physics, 
 Osaka University, Toyonaka, Osaka 560-0043, 
 Japan
}

\author{Osamu Seto}
 \email{seto@physics.umn.edu}
 \affiliation{
 Department of Architecture and Building Engineering,
 Hokkai-Gakuen University,
 Sapporo 062-8605, Japan
}

%

\begin{abstract}
%
We investigate thermal leptogenesis in a supersymmetric
 neutrinophilic Higgs model 
 by taking phenomenological constraints into account,
 where, in addition to
 the minimal supersymmetric standard model,     
 we introduce an extra Higgs field with
 a tiny vacuum expectation value (VEV) which  
 generates neutrino masses.  
Thanks to this tiny VEV of the neutrinophilic Higgs, 
 our model allows to reduce the mass of
 the lightest right-handed (s)neutrino
 to be ${\cal O}(10^5)$ GeV 
 as keeping sufficiently large CP asymmetry 
 in its decay.  
Therefore, the reheating temperature after inflation
 is not necessarily high, hence this 
 scenario is free from gravitino problem. 
%
\end{abstract}


\preprint{OU-HET 712/2011} 
\preprint{HGU-CAP 011} 

\vspace*{3cm}
\maketitle




The origin of cosmological baryon asymmetry is one of 
 the most important questions in both particle physics and cosmology.
Among various mechanisms of generating the suitable
 baryon asymmetry, 
 leptogenesis~\cite{FukugitaYanagida}
 is one of the most attractive scenarios. 
Particularly, thermal leptogenesis requires
 only thermal excitation of right-handed Majorana neutrinos which generate 
 tiny neutrino masses via a seesaw mechanism~\cite{Type1seesaw}, and 
 provides several implications for the spectrum~\cite{Buchmulleretal} of 
 light neutrino masses confirmed by neutrino oscillation
 experiments~\cite{Strumia:2006db, analysis}.  
However, a realization of thermal leptogenesis has a difficulty of 
 ``gravitino problem''~\cite{GravitinoProblem} 
 in supersymmetric models with R-parity. 
In order to avoid the overproduction of gravitinos,  
 the reheating temperature after inflation
 $T_R$ must not be so high  
 to thermalize right-handed (s)neutrinos~\cite{GravitinoProblem2}. 
Therefore, gravitino problem is a serious obstacle 
 in a usual Type-I seesaw~\cite{Type1seesaw}, where 
 tiny neutrino masses of order 0.1 eV 
 is obtained through superheavy
 right-handed neutrinos. 

How about an alternative idea, a neutrinophilic Higgs doublet
 model~\cite{Ma,Nandi,Ma:2006km,Davidson:2009ha,Logan:2010ag,HabaHirotsu}?
Here the smallness of neutrino masses originates from 
 a tiny VEV of neutrinophilic Higgs doublet, 
 and neutrino Yukawa couplings are not tiny anymore. 
Recently, we have shown that 
 thermal leptogenesis could work at a low energy scale 
 in a neutrinophilic Higgs doublet model  
 without 
 gravitino problem~\cite{HabaSeto}.
However, it is also worried that  
 enlarge neutrino Yukawa couplings might 
 give rise to sizable processes of lepton flavor violations (LFVs). 
Thus, in this paper, 
 we will show that thermal leptogenesis 
 surely works without gravitino problem 
 in a supersymmetric neutrinophilic Higgs doublet model 
 after carefully 
 taking other phenomenological constraints into account. \\


The supersymmetric neutrinophilic Higgs model has 
 a pair of neutrinophilic Higgs doublets
 $H_{\nu}$ and $H_{\nu'}$ in addition to 
 up- and down-type two Higgs doublets $H_u$ and $H_d$ in
 the minimal supersymmetric standard model (MSSM). 
A discrete $Z_2$-parity to discriminate
 $H_u (H_d)$ from $H_{\nu}(H_{\nu'})$ is also introduced,
 and its charges (and also lepton number) are assigned as the following table. 
\begin{table}[h]
\centering
\begin{center}
\begin{tabular}{|l|c|c|} \hline
fields  &  $Z_{2}$-parity & lepton number \\ \hline\hline
MSSM Higgs doublets, $H_u, H_d$  &  $+$ &  0 \\ \hline
new Higgs doublets, $H_{\nu}, H_{\nu'}$ 
 &  $-$ & 0 \\ \hline
right-handed neutrinos, $N$  &  $-$ & $1$ \\ \hline
others  &  $+$ & $\pm 1$: leptons, $0$: quarks \\ \hline
\end{tabular}
\end{center}
\end{table}
%
Under the discrete symmetry, 
 the superpotential is given by
\begin{eqnarray}
 W &=&y^{u}\bar{Q} H_u U_{R}
 +y^d \bar{Q} {H_d}D_{R}+y^{l}\bar{L}H_d E_{R} \nonumber \\
&& +y^{\nu}\bar{L}H_{\nu}N +\frac{1}{2}M {N^{}}^2 \nonumber \\
&& +\mu H_uH_d + \mu' H_\nu H_{\nu'} 
+\rho H_uH_{\nu'} + \rho' H_\nu H_d ,
\end{eqnarray}
 where we omit generation indexes. 
The $Z_2$-parity plays a crucial role of 
 suppressing 
 tree-level flavor changing neutral currents (FCNCs), 
 and  
 is assumed to be softly broken 
 by tiny parameters of $\rho$ and $\rho' (\ll \mu, \mu' )$.
We expect that 
 supersymmetry breaking soft squared masses can trigger 
 suitable electro-weak symmetry breaking. 
The Higgs potential is given by 
\begin{eqnarray}
 V &=& |\mu|^2(H_u^\dag H_u + H_d^\dag H_d ) + |\mu'|^2(H_{\nu}^\dag H_{\nu} + H_{\nu'}^\dag H_{\nu'} ) \nonumber \\
  && + \frac{g_1^2}{2} \left( H_u^\dag \frac{1}{2} H_u - H_d^\dag\frac{1}{2} H_d 
     + H_{\nu}^\dag \frac{1}{2} H_{\nu} - H_{\nu'}^\dag \frac{1}{2}H_{\nu'} \right)^2  \nonumber \\
  && + \sum_a \frac{g_2^2}{2} \left( H_u^\dag \frac{\tau^a}{2} H_u + H_d^\dag\frac{\tau^a}{2} H_d 
     + H_{\nu}^\dag \frac{\tau^a}{2} H_{\nu} + H_{\nu'}^\dag \frac{\tau^a}{2}H_{\nu'} \right)^2  \nonumber \\
  && + m_{H_u}^2 H_u^\dag H_u  + m_{H_d}^2 H_d^\dag H_d 
     + m_{H_\nu}^2 H_{\nu}^\dag H_{\nu}+ m_{H_{\nu'}}^2 H_{\nu'}^\dag H_{\nu'} \nonumber \\
  && + B \mu H_u \cdot H_d + B' \mu' H_{\nu}\cdot H_{\nu'}
 + \hat{B} \rho H_u \cdot H_{\nu'} + \hat{B}' \rho' H_{\nu}\cdot H_{d}
 + {\it h.c.} , 
\end{eqnarray}
 where we have omitted tiny $\rho^2$ and $\rho'^2$ mass terms.
$\tau^a$ and dot represent a generator of $SU(2)$ and
 its anti-symmetric product, respectively, and  
 $g_1 $ ($g_2$) is a 
 gauge coupling constant of $U(1)_Y $ ($SU(2)_L$). 
$m_{H_u}^2 (m_{H_d}^2, m_{H_\nu}^2, m_{H_{\nu'}}^2)$ and
 $B (B', \hat{B}, \hat{B}')$ are soft SUSY breaking parameters.
The tiny soft $Z_2$-breaking parameters, 
 $\rho, \rho' $,  
 generate  
 a large hierarchy of $v_{u,d} (\equiv \langle H_{u,d}\rangle) \gg 
 v_{\nu, \nu'}(\equiv \langle H_{\nu, \nu'}\rangle)$ through 
 stationary conditions,  
\begin{equation}
\left(
\begin{array}{cc}
 m_{H_{\nu}}^2 + \mu'^2 + \frac{m_Z^2 }{2}\frac{\tan^2\beta - 1 }{\tan^2\beta + 1 }   & -B_{\mu'} \mu' \\
 -B_{\mu'} \mu' &  m_{H_{\nu'}}^2 + \mu'^2 - \frac{m_Z^2}{2}\frac{\tan^2\beta - 1 }{\tan^2\beta + 1 } 
\end{array}
\right) 
\left(
\begin{array}{c}
 v_{\nu}  \\
 v_{\nu'}
\end{array}
\right)
 \simeq
\left(
\begin{array}{cc}
 - (\mu' \rho + \mu \rho')    & B_{\rho'} \rho' \\
 B_{\rho} \rho &  - (\mu \rho + \mu' \rho') 
\end{array}
\right) 
\left(
\begin{array}{c}
 v_u  \\
 v_d
\end{array}
\right) . 
\end{equation}
For example, 
 $v_{\nu}\sim 1$ GeV is obtained from 
 $\rho,\rho' \sim 1$ GeV with  
 Higgs mass parameters of     
 ${\cal O}(10^2)$ GeV.  
At the vacuum of $v_{\nu, \nu'}\ll v_{u,d}$ that we are interested in,
 physical Higgs bosons originated from 
 $H_{u, d}$
 are almost decoupled from
 those from $H_{\nu,\nu'}$. 
The former, $H_{u,d}$, almost constitute Higgs bosons in the MSSM;
 two CP-even Higgs boson $h$ and $H$, 
 one CP-odd Higgs boson $A$, and charged Higgs boson $H^\pm$, 
 while the latter, $H_{\nu, \nu'}$, 
 constitute 
 two CP-even Higgs bosons $H_{2,3}$,
 two CP-odd bosons $A_{2,3}$,
 and two charged Higgs bosons $H^\pm_{2,3}$. 
The two physical charged Higgs bosons are 
 given by 
\begin{equation}
\left(
\begin{array}{c}
 H_{\nu}^\pm  \\
 H_{\nu'}^\pm 
\end{array}
\right) =
\left(
\begin{array}{cc}
 \cos \alpha_c  & -\sin \alpha_c \\
 \sin \alpha_c & \cos \alpha_c
\end{array}
\right) 
\left(
\begin{array}{cc}
 H_2^\pm  \\
 H_3^\pm
\end{array}
\right), 
\end{equation}
where
%
$ \tan 2\alpha_c = 2 B_{\mu'} \mu'/\{ m_{H_{\nu}}^2 -
  m_{H_{\nu'}}^2 + (m_Z^2 - 2 m_W^2)\frac{\tan^2\beta - 1 }{\tan^2\beta
  + 1 }\}$
%
and
 $\tan\beta = v_u/v_d$.

Through the seesaw mechanism, masses of light neutrinos are given by 
\begin{equation}
m_{ij} = \sum_k \frac{y^{\nu}_{ik}v_{\nu} y^{\nu}{}^T_{kj}v_{\nu}}{M_k}. 
\label{66}
\end{equation}
For fixed right-handed neutrino masses,
 a tiny VEV of $v_{\nu}$ requires larger neutrino
 Yukawa couplings $y^{\nu}$ than 
 conventional seesaw scenarios. 
The neutrino masses may be also received 
 radiative corrections as 
 $m_{\nu}^{loop} \sim 
 -\lambda v_d^2  m_{H^\pm_{2,3}}^2/(8\pi^2 M)$,  
 where 
 we assume $M \gg m_{H^\pm_{2,3}}$, and 
 $\lambda$ is a coupling of
 one-loop induced scalar 
 interaction, 
 $\lambda (H_\nu \cdot H_d)^2$. 
Notice that 
 the radiative induced mass is smaller than
 the tree-level mass of Eq.~(\ref{66}) 
 as long as $\lambda<16\pi^2 v_\nu^2/v_d^2 $~\cite{Haba:2011nb}. 
Thus, 
 we can neglect radiative corrections 
 of neutrino masses, since   
 our model induces 
 $\lambda \sim g_2^4\rho'^2 /(32 \pi^2 m_{\tilde{\chi}^\pm}^2)  \sim 10^{-10}$,
 where $m_{\tilde{\chi}^\pm}$ is a chargino mass. 
For this estimation, 
 we have used   
 $v_\nu/v_d\sim 10^{-2}$ which will be a suitable 
 parameter region in the following discussions.  
Actually, 
 there are two 1-loop diagrams which contribute  
 $\lambda$, but the chargino 1-loop diagram 
 dominates a Higgs 1-loop diagram, and we neglect the latter.  
Notice that 
 (s)top 1-loop diagram is negligible 
 due to the $Z_2$-parity. 
Anyhow, the value of $\lambda$ is tiny, since 
 it is not induced until
 1-loop diagrams including both $Z_2$- and SUSY-breaking
 effects. 
The neutrino mass matrix in Eq.~(\ref{66}) can reproduce neutrino oscillation 
 experiments, and we will use a concrete 
 value of $\Delta m^2_{\rm atm}$~ \cite{atm} 
 in the following analyses. \\

Now, let us discuss thermal leptogenesis in this model. 
A resultant baryon asymmetry generated via thermal leptogenesis 
 is generally given by 
\begin{equation}
\frac{n_b}{s} \simeq  C \kappa \frac{\varepsilon}{g_*}  , 
 \label{b-sRatio}
\end{equation}
where $\left. g_*\right|_{T=M_1} = {\cal O}(100)$
 is the effective degrees of freedom of
 relativistic particles in thermal bath, 
 and $\varepsilon$ is the total CP asymmetry of right-handed
 (s)neutrino decay. 
Dilution (or efficiency) factor 
 $ \kappa \leq {\cal O}(0.1) $ denotes
 the dilution by washout processes, and
 the coefficient $C$ is a factor of
 the conversion from lepton to baryon asymmetry
 by the sphaleron~\cite{KRS}.  
The decay rate of the lightest right-handed neutrino $N_1$ into left-handed (s)leptons
 and $H_{\nu} (\tilde{H}_{\nu})$-like Higgs bosons (higgsinos) $\Gamma_{N_1}$ and
 that of the lightest right-handed
 sneutrino $\tilde{N}_1$
 into left-handed sleptons (leptons)
 and $H_{\nu} (\tilde{H}_{\nu})$-like Higgs bosons (higgsinos) $\Gamma_{\tilde{N}_1}$
 are given by
\begin{eqnarray}
\Gamma_{N_1} = \Gamma_{\tilde{N}_1}= \sum_j\frac{y^{\nu}_{1j}{}^\dagger y^{\nu}_{j1}}{4\pi}M_1 = \frac{(y^{\nu}{}^\dagger y^{\nu})_{11}}{4\pi}M_1.
\end{eqnarray}
We would note here that 
 physical mass eigenstate of
 $H_{\nu} (\tilde{H}_{\nu})$-like Higgs boson (higgsino)
 has tiny component of 
 $H_{u} (\tilde{H}_{u})$ through the tiny $Z_2$-breaking parameter $\rho$. 
The condition for out of equilibrium in decay of right-handed (s)neutrino
 $\Gamma_{N_1 (\tilde{N}_1)} < H|_{T=M_1}$
 requires that 
 the lightest left-handed neutrino is almost 
 massless $m_1 \simeq 0$ and $y^{\nu}_{i1}$ are very small,  
 where $H$ is the Hubble parameter and $T$
 is the temperature of radiation. 
For the neutrino Yukawa couplings of 
 $y^{\nu}_{i1} \ll y^{\nu}_{i2}, y^{\nu}_{i3}$ and
 hierarchical right-handed neutrino mass spectrum~\cite{SUSYFandG}, 
 the total CP asymmetry of right-handed (s)neutrino
 decay is given by 
\begin{eqnarray}
\varepsilon &\equiv& \varepsilon(N \to l H) +\varepsilon(N \to \tilde{L}\tilde{H}) 
                          +\varepsilon(\tilde{N} \to l H)+\varepsilon(\tilde{N} \to \tilde{L} H) \nonumber \\
& \simeq & -\frac{3}{16\pi} 10^{-5} \left(\frac{0.1 {\rm GeV}}{v_{\nu}}\right)^2
  \left(\frac{M_1}{10^3 {\rm GeV}}\right)
  \left(\frac{m_{\nu}}{0.05 {\rm eV}}\right) \sin\theta . 
\label{CPasym}
\end{eqnarray}
Here $\theta$ is an effective CP violating phase, which 
 is significantly enhanced due to 
 the tiny $v_{\nu}$.
In order to obtain the observed baryon asymmetry 
 in our Universe $n_b/s \simeq 10^{-10}$~\cite{WMAP}, 
 $\varepsilon \gtrsim 10^{-7}$ is required.
For the conventional Type-I 
 seesaw in the MSSM with superheavy right-handed neutrinos  
 (where the neutrino Dirac mass term is generated through $v_u$), 
 $\varepsilon \gtrsim 10^{-7}$ 
 means $M_1 \gtrsim 10^9$ GeV,
 which is so-called Davidson-Ibarra bound for models with
 hierarchical right-handed neutrino
 mass spectrum~\cite{LowerBound,Davidson:2002qv}. 
In contrast, $v_u$ is replaced by $v_{\nu} (\ll v_u)$ in our model, 
 and as the result, an enough large $\varepsilon$ can be obtained
 even for a smaller $M_1$ 
 than that derived by Davidson-Ibarra bound.

Lepton number violating scatterings act as washout processes of 
 the generated lepton number asymmetry.
Those scatterings are classified into two classes;
 the lepton number is violated by one ($\Delta L =1$) and
 by two ($\Delta L = 2$). 
The $\Delta L =1$ scattering rates are proportional to $\Gamma_{N_1}$ and
 hence can be minimized by its appropriate choice.
We should notice that 
 $\Delta L=1$ processes such as 
 $LN\to H\to Q\bar{t}$ are negligible, 
 since 
 the mixing between $H_\nu (\tilde{H}_\nu)$
 and $H_u (\tilde{H}_u)$, $H_d (\tilde{H}_d)$ 
 is negligible due to tiny $Z_2$ breaking. 
On the other hand, the $\Delta L =2$ scatterings are potentially
 dangerous, and 
 relevant scatterings for the MSSM have been
 studied in Ref.~\cite{Plumacher:1997ru}.
The decoupling condition for $\Delta L =2$ lepton
 number violating scatterings 
 $\gamma_A^{\Delta L}$ in Ref.~\cite{Plumacher:1997ru} 
 is applicable to our model, which is roughly estimated as 
\begin{eqnarray}
 \sum_i \left(\sum_j \frac{  y^{\nu}_{ij} y^{\nu}_{ji}{}^{\dagger} v_{\nu}^2}{M_j}\right)^2
   <  \pi^3 \zeta(3) \sqrt{\frac{\pi^2 g_*}{90}} \frac{v_{\nu}^4}{T M_P} , 
 \label{L2DecouplingCondition}  
\end{eqnarray}
 for $T < M_1$.
Other scattering processes also give similar conditions.
For a lower $v_{\nu}$, washout processes are more significant.
Inequality~(\ref{L2DecouplingCondition}) gives
 the lower bound on $v_{\nu}$ to avoid too strong washout.

As we have shown above, 
 a sufficient CP violation $\varepsilon = {\cal O}(10^{-6})$ can be realized 
 for $v_{\nu} = {\cal O}(1)$ GeV 
 in the hierarchical right-handed
 neutrino with $M_1$ of ${\cal O}(10^5 - 10^6)$ GeV.
This implies
 that the reheating temperature after
 inflation $T_R$ of ${\cal O}(10^6)$ GeV is high enough 
 to produce right-handed neutrinos by thermal scatterings.
Thus, this class of model with $v_{\nu} = {\cal O}(1)$ GeV is 
 a solution to compatible with thermal leptogenesis
 in gravity mediated supersymmetry breaking with unstable gravitino.

\begin{figure}[t] 
\begin{center}
  \epsfig{file=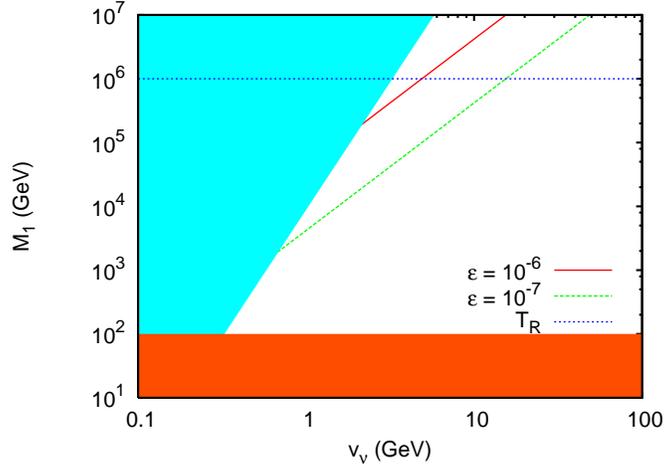,width=9cm}
\end{center}
 \caption{Available region for leptogenesis.
 In the red brown region, 
 the out of equilibrium decay of the lightest right-handed neutrino
 is not possible.
 In turquoise region, $\Delta L=2$ washout effect is too strong.
 The red and green line are contours for the CP asymmetry. 
 The blue line represents a typical value of upper bound on $T_R$
 to avoid gravitino problem.
}
\label{fig:AvailableRegion}
\end{figure}
%


\vspace{5mm}

Next, 
 let us investigate phenomenological constraints in our model.
The most severe constraint comes from LFV decay processes,
 particularly, $\mu \to e \gamma$.  
There are LFV processes from 1-loop processes triggered by
 $y^{\nu}$ 
 (loops of $N$-$H^\pm$ and $\tilde{N}$-$\tilde{\chi}^\pm$) 
 in addition to the MSSM processes. 
A branching ratio of the LFV is given by
\begin{eqnarray}
 && B(l_{\alpha} \rightarrow l_{\beta} \gamma ) \nonumber \\
 &=& \frac{3 \alpha_{\rm em}}{64 \pi G_F^2} 
 \left| \sum_i  y_{\alpha i} y^{\dagger}_{i \beta} \left\{ \frac{\cos^2\alpha_c}{M_{H_2^\pm}^2} 
 F\left(\frac{M_i^2}{M_{H_2^\pm}^2}\right)
 + \frac{ \sin^2\alpha_c }{M_{H_3^\pm}^2}
 F\left(\frac{M_i^2}{M_{H_3^\pm}^2}\right)
 - \frac{M_{\tilde{\chi}^\pm}}{2 m_{\mu} M_{\tilde{N}_i}^2} 
 G\left(\frac{M_{\tilde{\chi}^\pm}^2}{M_{\tilde{N}_i}^2}\right) \right\}
 + {\rm MSSM \, processes} \right|^2 
\end{eqnarray}  
 where 
\begin{eqnarray}
 && F(x) = \frac{1}{6(1-x)^4}(1 -6 x +3 x^2 + 2 x^3 -6 x^2 \ln x), \\
 && G(x) = \frac{1}{(1-x)^3}(-3 + 4 x - x^2 -2 \ln x). 
\end{eqnarray} 
Here $G_F$ is the Fermi coupling constant.  
$M_{\tilde{N}_i}$ is the mass of $N_i$, and 
$M_{\tilde{\chi}^\pm}$ is the mass of
 $\tilde{H}_{\nu} (\tilde{H}_{\nu'})$-like chargino.
In all parameter region except for $M_i^2 \approx H_{2, 3}^\pm$,
 the chargino-loop contribution
 involving $G( M_{\tilde{\chi}^\pm}^2/M_{\tilde{N}_i}^2 )$ is dominant and 
 the charged Higgs boson loop contribution
 depending upon $F( M_i^2/M_{H_{2, 3}^\pm}^2 )$ is negligible.
Thus, almost independent from 
 the charged Higgs boson masses $H_{2,3}^\pm$ and its mixing angle 
 $\alpha_c$, 
 the additional contribution to the LFV decay is estimated as
 $B(l_{\alpha} \rightarrow l_{\beta} \gamma ) \lesssim 10^{-13}$, 
 which is much smaller than the current experimental bounds, 
 $B(\mu \rightarrow e \gamma ) < 1.2 \times 10^{-11}$,
 $B(\tau \rightarrow e \gamma ) <  3.3 \times 10^{-8}$, and
 $B(\tau \rightarrow \mu \gamma ) < 4.4 \times 10^{-8}$~\cite{PDG}. 
We here take 
 a parameter region where leptogenesis
 effectively works without 
 washout effects.
The similar diagram (initial and final states are both muon) 
 induces
 a deviation of muon anomalous magnetic moment
 $a_{\mu} \equiv (g_{\mu}-2)/2$.
Similarly, additional contributions
 to $a_{\mu}$ from above loop processes is turned out to be 
 $\Delta a_\mu = {\cal O}(- 10^{-15})$, which is sufficiently tiny. 


We here summarize all conditions for successful thermal leptogenesis, 
 and the result is presented in the Figure~\ref{fig:AvailableRegion}.
The horizontal axis is the VEV of neutrino Higgs $v_{\nu}$ 
 and the vertical axis is the mass of the lightest right-handed
 neutrino $M_1$ in hierarchical right-handed neutrino mass spectrum.
In the brown region, the lightest right-handed neutrino decay
 into $H_{\nu}$-like Higgs boson and lepton
 is kinematically not allowed.
In turquoise region corresponds to inequality
 of Eq.(\ref{L2DecouplingCondition}), where 
 $\Delta L=2$ washout effect is too strong.
The red and green line are the contours
 of the CP asymmetry of $\varepsilon=10^{-6}$ and $10^{-7}$,
 respectively.
Thus, in the parameter region near 
 above the line of $\varepsilon = 10^{-7}$, 
 thermal leptogenesis easily works 
 even with hierarchical masses of right-handed neutrinos. \\

We have investigated thermal leptogenesis in a supersymmetric 
 neutrinophilic Higgs doublet model with 
 taking account of phenomenological constraints and gravitino 
 problem.  
One of the attraction of neutrinophilic Higgs models is
 that the neutrino Yukawa couplings are not necessarily tiny anymore. 
They can enhance the CP asymmetry of right-handed (s)neutrino decay, 
 however might also enhance
 the washout rate of generated lepton asymmetry
 and magnitudes of LFV processes, 
 similtaneously. 
We have found that 
 the suitable baryon asymmetry is reproduced 
 with the suitable neutrino masses of 
 ${\cal O}(10^{-1})$ eV,   
 in which expected LFVs are consistent with current experiments 
 and the strong $\Delta L = 2$ washout can be avoided. 
To generate and thermalize relatively 
 light right-handed neutrino 
 with mass of ${\cal O}(10^5)$ GeV, 
 the reheating temperature 
 is low enogh to avoid gravitino problem. 
 
At the end, 
 we comment on gauge coupling unification (GCU). 
It can be achieved by introducing extra vector-like 
 $SU(3)_c$-triplet particles, $d, \bar{d}$. 
We also introduce an additional $Z_2$-parity,
 and make only $d, \bar{d}$ have odd-charge of it. 
Therefore, $d, \bar{d}$ have no Yukawa interactions 
 with ordinal quarks and leptons as possessing their
 heavy masses of $W \sim \mu' d \bar{d}$.   
This field content is similar to
 so-called Nelson-Barr model~\cite{NB}
 and its supersymmetric version proposed in Ref.~\cite{Dine:1993qm}. 
Thus, our model could solve the strong CP problem and achieve the 
 suitable GCU as well as 
 realize thermal leptogenesis without gravitino problem. \\

%
We are grateful to Y.~Okada for valuable suggestions. 
This work is partially supported by Scientific Grant by Ministry of 
 Education and Science, Nos. 20540272, 20039006, 20025004 (N.H.), and
 the scientific research grants from Hokkai-Gakuen (O.S.). 



\end{document}